\documentstyle[preprint,aps]{revtex}

\newcommand{\dslash}{i \partial \hspace{-.5em}/\hspace{.2em}}

\newcommand{\Aslash}{A \hspace{-.5em}/\hspace{.2em}}

\newcommand{\beq}{\begin{eqnarray}}
\newcommand{\enq}{\end{eqnarray}}
\newcommand{\beqn}{\begin{eqnarray*}}
\newcommand{\enqn}{\end{eqnarray*}}
\newcommand{\rar}{\rightarrow}
\newcommand{\ral}{\leftrightarrow}
\newcommand{\non}{\nonumber}

\newcommand{\der}{\partial}

\def\ie {{\it i.e., }}
\def\eg {{\it e.g., }}

\def\bitem{\begin{itemize}}
\def\eitem{\end{itemize}}
\newcommand{\li}{[{\lambda^i}]}

\newcommand{\lk}{[{\lambda^k}]}
\newcommand{\rf}[1]{(\ref{#1})}

\newcommand\N{N_c}

\def\ie {{\it i.e., }}
\def\eg {{\it e.g., }}

\begin{document}
\draft
\preprint{\vbox{
\hbox{SUNY-NTG-95-33 }
\hbox{hep-ph/9510326 } }}

\title{Heavy Hadrons and QCD Instantons}

\author{S. Chernyshev, M.A. Nowak\footnote{GSI Darmstadt, Postfach 110552,
D-64220 Darmstadt, Germany and Institute of Physics, Jagellonian University,
PL-30059 Cracow, Poland} and I. Zahed}

\address{Department of Physics, SUNY at Stony Brook,
Stony Brook, New York 11794}
\bigskip
\date{\today}
\maketitle
\begin{abstract}
Heavy hadrons are analyzed in a random and dilute gas of instantons.
We derive the instanton-induced interactions between
heavy and light quarks at next to leading order in the heavy quark mass and
in the planar approximation, and
 discuss their effects on the hadronic spectrum.
The role of these interactions in the formation of exotic hadrons is also
discussed.
\end{abstract}
\vskip 1cm

\pacs{PACS numbers: 11.15.Pg, 12.38.Cy, 12.38.Lg, 12.39.Hg, 14.20.Lq.}

\narrowtext

\section{Introduction}

Hadrons with one or many heavy quarks exhibit a new type of symmetry:
invariance under spin-flip of the heavy quark \cite{isgwis}. This
invariance can
be used to organize the hadronic structure and properties of heavy-light
systems. A number of relations follow both in the spectrum and among form
factors of heavy hadrons when the mass of the heavy quark is taken to infinity
\cite{hqet}.

The interplay between
 light and heavy degrees of freedom in heavy-light hadrons
can be clarified in the heavy-quark limit by combining chiral symmetry with
heavy quark symmetry \cite{wise}. The basic observation is to note that the
hard part in the heavy quark field is kinematical and factorizable. The
remaining part is soft and constrained by chiral dynamics. Hence, the soft
physics in heavy-light systems can be analyzed in a way similar to
the light-light systems. A qualitative understanding of this part can be
achieved by using QCD inspired models.

In this paper we discuss the effects of a random gas of instantons
and antiinstantons on mesons and baryons containing one or several
heavy quarks. We analyze the correlation functions of various hadrons with one
or many heavy quarks in inverse powers of the heavy quark mass $m_Q$ using a
succession of Foldy-Wouthuysen transformations prior to radiative corrections.
In section 2, we give a brief
 summary of the salient properties of some typical
heavy-light systems. In section 3, we show how the heavy meson correlator
may be systematically analyzed in inverse powers of the heavy quark mass, in
the planar approximation. Recoil and magnetic corrections to both the heavy
quark propagator and the heavy-meson correlator are evaluated at next to
leading order in the heavy quark mass. These results
 are quantified in the form
of effective interactions between heavy and light constituent quarks. In
section 4, we estimate the effects of the induced effective interactions and
heavy meson and baryon spectra. The results are in overall agreement with
the constituent quark model and heavy solitons. We briefly discuss
similar effects in exotic hadronic configurations. Our conclusions are
summarized in section 5.

\section{Generalities}

Throughout, heavy hadrons will be understood as mesons or baryons
with at least one heavy quark. Although we will be interested in taking the
heavy quark mass to infinity and then relaxing it, we will in fact specifically
have in mind for a heavy quark,
the bottom quark ${\bf b}$ with a mass of $4.7-5.3$ GeV,
the charmed quark ${\bf c}$ with a mass of $1.3-1.7$ GeV,
and to some extent the strange quark ${\bf s}$ with a mass of $100-300$ MeV.

Heavy mesons $Q\overline q$ may be organized in their ground state into
multiplets with $I(J^P)=\frac 12 (0^-,1^-)$. In the heavy quark limit, the
multiplets are invariant representations of the heavy quark symmetry group
(essentially left spin rotation). Empirically $(K, K^*) = (493,892)$ MeV,
$(D,D^*) =(1869,2010)$ MeV and $(B,B^*)=(5278, 5324)$ MeV. $D^*$ decay is
mainly through $D\pi$.

The heavy baryons with one heavy quark will be of the type $Qqq$.
The conventions being ${\bf \Lambda}_Q = 0\frac 12^+$,
${\bf \Sigma}_Q = 1\frac 12^+$ and ${\bf \Sigma}_Q^* = 1\frac 32^+$.
The ones with two heavy quarks will be of the type $QQq$. Few heavy baryons
have already been observed. Organizing them in multiplets invariant under
heavy quark symmetry, we have
$\Lambda_s, (\Sigma_s,\Sigma_s^*) = 1116, (1195, 1385)$ MeV,
$\Lambda_c, (\Sigma_c, \Sigma_c^*) = 2284, (2455, 2530\pm 5 \pm
 5\cite{BUBBLE}\footnote{
To be confirmed by other experiments.})$ MeV, and
$\Lambda_b, (\Sigma_b, \Sigma_b^*) = 5641, (..., ...)$ MeV. The dots
refer to yet to be measured masses. Other measured heavy (charmed) baryons
include $(\Xi_c,\Xi_c^*)= (2468, 2642.8\pm 2.2$ \cite{CLEOII}$)$ MeV
and $\Omega_c=2704$ MeV.   Heavy baryons with more than one heavy
quark have not been found yet.

The basic principles at work in a heavy light system are best illustrated
using a simple bag model description. If we were to insert a heavy source
in a spherical cavity of radius $R$, then the total
energy can be organized using the bare heavy quark mass $m_Q$ following
$E=m_Q + E_0 m_Q^0 + E_1 m_Q^{-1} + ...$. The contribution $E_0$ refers to the
energy of the light quarks present in the cavity and is standard \cite{bag}.
The contribution $E_1$ corresponds to

\beq
\frac{E_1}{m_Q} = \frac{\big ( \pi /R \big )^2}{2m_Q}
+ \frac{{\vec\mu}^a\cdot {\vec B}^a}{2m_Q}
\label{A1}
\enq
where the first term is the recoil of the heavy quark, and the second term
is the magnetic interaction between the average magnetic field induced by the
light quark at the center of the bag, and the magnetic moment of the heavy
quark.
While schematic, (\ref{A1}) captures the essence  of the $1/m_Q$ corrections
in heavy quark physics. Using standard bag model parameters \cite{bag}, we
have for charmed mesons (recoil, spin) $\sim$ (400, 20) MeV, while for bottom
mesons (recoil, spin) $\sim$ (100, 5) MeV \cite{zah}.

\section{Heavy Hadrons in an Instanton Gas}

In what follows, we will try to understand the origins of $E_0$ and $E_1$
from a microscopic description of the QCD vacuum using a random gas of
instantons and antiinstantons.

\subsection{Heavy Quark Expansion}

Consider the correlation function of a heavy-light meson in the QCD vacuum.
In Minkowski space, it reads

\beq
{\cal C}^{\pm}_{\Gamma}(x,x') = \langle \, T \left (
\overline{q} \Gamma_{\pm} \psi (x) \overline{\psi} \Gamma_{\pm} q(x')
\right ) \, \rangle
\label{1}
\enq
with $\Gamma_{\pm} = ({\bf 1}, \gamma)\times (1\pm \gamma^0)/2
\times ({\bf 1} , T)$ a non-relativistic source with arbitrary flavor.
Here $q(x)$ refers to the light quark, and $\psi (x)$ to the heavy quark
with bare mass $m_Q$. For $m_Q$ much larger than the typical scale of the
problem $\Lambda_{\rm QCD}$, one may use $\Lambda_{\rm QCD}/m_Q$ expansion
to analyze \rf{1}.
We perform this expansion using  a Foldy-Wouthuysen
 transformation of the heavy quark field \cite{kov}.
\beq
\psi (x) \sim e^{-i\gamma_0 m_Q t}\,\,
e^{-i\sigma_{0i}[\nabla^0, \nabla^i]/4m_Q^2}\,\,
e^{-i\vec\gamma\cdot\vec\nabla/2m_Q}\,\,Q (x)
\label{fwout}
\enq
where $\nabla =\partial - i A$ and with $g$ (the gauge coupling) set to one.
The first transformation  rescales the momenta, the second
eliminates the odd parts, and the third removes the mass term. The successive
transformations in (2) are vector-like, unitary and gauge-covariant.
 In terms of (\ref{fwout}) the QCD part of the
action for the heavy field  $\psi$ becomes
\beq
{\cal L}_{\psi} \sim \overline{Q}  i\gamma^0\nabla^0 Q  +
\overline{Q}\hat{O}_1 Q + {\cal O}(m_Q^{-2})
\label{3c}
\enq
where operator $\hat{O}_1$ is defined as
\beq
 \hat{O}_1 =-\frac{\vec\nabla^2}{2m_Q} -\frac {\vec\sigma\cdot \vec B}
{2m_Q}
\label{O1}
\enq
 with $B^i= -i\epsilon^{ijk} [\nabla^i, \nabla^j]$. Equation
(\ref{3c}) has the expected
FW form to order $m_Q^{-1}$. In the first term in (\ref{O1}) we
recognize the recoil,
in the  second we recognize spin effect on the heavy quark in
external field . Under (\ref{fwout}) the heavy meson source shifts
\beq
\overline{q}\Gamma \psi \sim \overline{q}\Gamma e^{-i\gamma^0 m_Q t}
\left( 1- \frac{i\vec\gamma\cdot\vec\nabla}{2m_Q}\right) Q
\label{4}
\enq
As a result, the correlator (\ref{1}) takes the generic form
\beq
{\cal C}_{\Gamma} \sim
\langle 0| \overline{q} \Gamma e^{-i\gamma^0 m_Q (t-t')} Q \overline{Q} \Gamma
q
|0 \rangle +
\langle 0| \overline{q} \Gamma e^{-i\gamma^0 m_Q (t+t')}
\bigg [ Q\overline{Q},
\frac{i\vec\gamma\cdot\stackrel{\leftarrow}{\nabla}}{2m_Q}\bigg ] \Gamma q
|0\rangle
\label{5}
\enq
 Mixing between
the particle and the antiparticle content of the correlator  (\ref{5})
drops out in the nonrelativistic limit \cite{cnz}.
Thus, in Euclidean space
\beq
{\cal C}^{\pm}_{\Gamma}(x,x')\sim - e^{\mp m_Q (\tau -\tau')}
\langle 0 | {\rm Tr} \left(\Gamma_{\pm} \,S_Q(x,x')\,\Gamma_{\pm} \, S(x' , x
)\right) |0 \rangle
\label{6}
\enq
where $S$ is the propagator of the light quark, and $S_Q$ is the
heavy quark propagator,
\beq
S_Q \sim S_{\infty} +S_{\infty} \hat{O}_1
 S_{\infty}  + {\cal O}(m_Q^{-2})
\label{A11}
\enq
with $S_{\infty}=\gamma_4/i\nabla_4$ being the free part.
 This construction can be carried out to
arbitrary orders in $1/m_Q$ \cite{kov,cnz}, given that the heavy
quark expansion is not upset by renormalization \cite{kov}.

\subsection{Heavy Quark Propagator}

The heavy quark propagator (\ref{A11}) may be analyzed in a random instanton
gas. In the planar approximation, the infinitely heavy quark propagator
satisfies the integral equation \cite{cnz,diapob}

\beq
S_{\infty}^{-1} = S_*^{-1} - \sum_{I, \overline I} \langle\left(
\,\,\Aslash_{4,I}^{-1} -S_{\infty} \right)^{-1}\rangle
\label{nlo}
\enq
where $\Aslash_4 =\gamma^4 A^4$ and $S_* =i\gamma^4\partial^4$.
The sum is over all instantons and anti-instantons and the averaging is
over the position $z_I$ and the $SU(N_c)$ color
orientation $U_I$, with
\beq
A_I(x)= U_I \; A(x-z_I, \rho) \; U_I^\dagger
\enq
Generically
\beq
\sum_{I,\overline I} \rightarrow \frac N2 \,\,\left(\frac 1{V_4}\int
d^4z_I\right) \,\,\int dU_I  + \left( I\rightarrow \overline I \right)
\sim \frac N{2V_4\N} \int d^4z_I \,\,{\rm Tr}_C
\left( I +\overline I \right)
\label{sum}
\enq
where ${\rm Tr}_C$ stands for a trace in  color space.
For a low instanton density $n=N/V_4\sim 1$ fm$^{-4}$, we can iterate
(\ref{nlo}) in powers of $n$. The result is

\beq
S_{\infty}^{-1} = S_*^{-1} + in \, \bigg ( \Theta_0 +
\frac 1{\rho m_Q} \Theta_1 + {\cal O} \big ( \frac 1{\rho^2 m_Q^2} \big )
\bigg ) + {\cal O} \big ( n^2 \big ) \label{den}
\enq
where the diluteness factor
is given by the dimensionless combination $n\rho^4\sim 10^{-3}$.
Substituting \rf{den} into (\ref{nlo}), we obtain

\beq
\Theta_0 = \int d^4z_I \,\,{\rm Tr}_c \left( S_*^{-1}
\bigg ( \frac 1{i\gamma^4\nabla_{4,I}}-S_* \bigg ) S_*^{-1} +
 I\rightarrow\overline I \right) \label{a}
\enq
and
\beq
\Theta_1 = \int d^4z_I \,\,{\rm Tr}_C \left( S_*^{-1}
\bigg ( \frac 1{i\gamma^4\nabla_{4,I}}-S_* \bigg ) {\cal O}_1
\bigg ( \frac 1{i\gamma^4\nabla_{4,I}}-S_* \bigg ) S_*^{-1} +
 I\rar \overline I \right) \label{b}
\enq
where ${\cal O}_1 = -\vec\nabla^2/2 - \vec\sigma\cdot \vec B/2$.
Both $\Theta_0$ and $\Theta_1$ are $\tau-$dependent. To proceed further, we
note that in coordinate space, the heavy quark propagator in the one instanton
background reads

\beq
<x|\frac 1{i\nabla_{4,I}} |0> =\delta (\vec x )\,\,\theta(\tau )\;
\frac {1 + \gamma^4}{2} \;
{\bf P}e^{i\int_0^{\tau}\, dsA_4(x_s-z_I )}
\label{prop}
\enq
with $x_s = (s,\vec x )$.
Inserting (\ref{prop}) into (\ref{a},\ref{b}) and
using the one-instanton configuration
\beq
A_{\mu}^a (x) = +{\overline\eta}^a_{\mu\nu} x_{\nu}
\left( \frac 1{x^2} - \frac 1{x^2 +\rho^2 }\right)
\enq
yield for large times
\beq
<x_{-\infty} | (\Theta_0 , \Theta_1 ) | x_{+\infty} > \sim
8\pi \rho^3 ( -4{\bf I}_0 , + {\bf I}_1 )
\label{mass}
\enq
with
\beq
({\bf I}_0 , {\bf I}_1 ) =
\int_0^{\infty} x^2 dx \;
\left\{ {\rm cos}^2 \big ( f_x /2 \big ) \,\, ,\,\,
\left( \; [\der_\mu \cos f_x ]^2 -
[\der_\mu \sin f_x ]^2 + (\rho A^a_\mu)^2 \cos 2 f_x  \; \right ) \right\}
\enq
where $f_x= \pi |x| /\sqrt{1+x^2}$.
There is no contribution to $\Theta_1$ from the spin part
$- \vec\sigma\cdot \vec B$. From \rf{mass} it follows that
the instanton induced shift in the heavy quark mass is
$\Delta_0 M_Q \sim 70\, $MeV \cite{diapob} from $\Theta_0$ and
$\Delta_1 M_Q \simeq 16 \, $MeV from $\Theta_1$
for a $c-$quark with $m_c = 1350\, $MeV. The recoil effect $\Delta_1 M_Q$
is an order of magnitude down compared to the naive bag estimate (\ref{A1}).
This can be understood by noting that in the presence of instantons, the
energy of a heavy quark can be rewritten schematically as

\beq
E = 32\pi \times n\rho^4 \times \bigg(  1/{\rho} + \frac {1/\rho^2}{m_Q} +
...\bigg)
\label{sche}
\enq
where the factors follow from (\ref{den}) and (\ref{mass}). (\ref{sche}) is
the analog of (\ref{A1}). For the instanton parameters used, and a charmed
quark, (\ref{sche}) yields

\beq
E \sim 32\pi \times 10^{-3}\times \bigg( 600 + \frac 12 \, 600 + ...\bigg) \sim
\bigg( 60 + 30 + ...\bigg) \,\, {\rm MeV}
\enq
which shows that the zeroth order shift in the mass is about 60 MeV, while
the recoil effect is about 30 MeV, as expected.

Similar arguments can be used for the light quark propagator.
The result is $S^{-1}\sim S^{-1}_0 + i\sqrt{n} \Sigma(x)$ with an average
light quark mass shift $\Delta M_q \sim 420$MeV \cite{cnz,pob}.

\subsection{Heavy Quark Correlator}

The correlator (\ref{1}), written generically as
${\cal C} \sim \langle S \otimes S_{\infty} \rangle$, receives contributions
from both planar and non-planar graphs and is usually hard to analyze in the
random gas approximation exactly. In the planar approximation, things
simplify. After resummation, the inverse correlator (\ref{1}) reads

\beq
{\cal C}^{-1} \sim S^{-1} \otimes S_{\infty}^{T\,-1}  - \sum_{I,\overline{I}}
\; \langle \,\,({S} -\Aslash_{I}^{-1}  )^{-1} \otimes
( {S}_{\infty} -\Aslash_{4,I}^{-1} )^{T\, -1}
 \,\,\rangle
\label{8}
\enq
The upper script $T$ is short for transpose. Here the
spin-flavor-color indices are left uncontracted and
the space-time indices are omitted.
Standard perturbation techniques in the massless sector gives

\beq
\frac 1{\Aslash_I^{-1} -S} =
\sum_n \; S^{-1}  \frac{ | \Phi_n \rangle \langle \Phi_n | }
{\langle \Phi_n | \, S^{-1} - \Aslash_I | \Phi_n \rangle }
\Aslash_I \label{pert}
\enq
where $|\Phi_n\rangle$
is the normalized eigenstate of the Dirac operator in a one-instanton
background,

\beq
\left( \dslash - \Aslash_I\right) \Phi_n (x-z_I ) =
\lambda_n \Phi_n (x-z_I )
\enq
It follows that the 't Hooft zero mode
$\Phi_0 (x-z_I )$ \cite{hoo}

\beq
\Phi_0 (x)= \frac 1\pi \frac{\rho}{(x^2 +\rho^2)^{3/2}}
\frac{x\cdot \gamma}{\sqrt{x^2}}
\left( \begin{array}{r}
        1 \\ -1  \end{array} \right) {\bf \varphi} \label{zero}
\enq
is dominant in (\ref{pert}). Thus, to leading order we obtain
\beq
{\cal C}^{-1} \sim
  S^{-1}  \otimes S_{\infty}^{T,\,-1} - n \int d^4 z_I \; {\rm Tr}_c
\left( [ \; L \; ]_I \; \otimes \; [ \; H \; ]_I + \,\,I\rightarrow{\overline
I}\right)
\label{12}
\enq
with
\beq
\label{13}
[ \; L \; ]_I &=& S_0^{-1}\left(\,\, \frac{| \Phi_0 \rangle \langle \Phi_0 |}
{i\sqrt{n}\Sigma_0} -S_0\,\,\right) S_0^{-1} \\[1pt]
[ \; H \; ]_I &=& S_*^{-1} \left( \frac 1{i\gamma^4\nabla_{4,I}}-S_*\right)
S_*^{-1}  +  
\frac 1{m_Q} \; S_*^{-1}
\left( \frac 1{i\gamma^4\nabla_{4,I}}-S_*\right) {\cal O}_1
\left( \frac 1{i\gamma^4\nabla_{4,I}}-S_*\right) S_*^{-1} \non
\enq
where $\Sigma_0  = \langle \Phi_0 | \Sigma | \Phi_0 \rangle
\sim (240 \; {\rm MeV})^{-1}$.

\subsection{Effective Interactions}

The inverse correlator (\ref{12})
allows for an immediate translation to effective interactions.
In the long wavelength
limit, the instanton size is small, and a local interaction between
the effective fields ${\bf Q}$ and ${\bf q}$ can be derived much like
the 't Hooft interaction between the light effective fields ${\bf q}$
\cite{shi,now}.
{}From the Bethe-Salpeter equation associated to (\ref{12}),
we read the vertex
\begin{equation}
\Gamma^{ac}_{bd} ( x,y, \; x^\prime , y^\prime ) =
- i n N_c \int d^4 z_I \int dU_I \,
\left ( U_i^a \, \langle x | [ \; L_I \; ]_j^i
| x^\prime \rangle {U^\dagger}_b^j \otimes U_k^c \langle y | [\; H_I \; ]^k_l
| y^\prime \rangle {U^\dagger}_d^l
+ I \rar \overline{I} \right ) \label{22}
\end{equation}
where the color matrices have been explicitly displayed.
This vertex function gives rise to an effective action ${\cal S}_I$
\begin{equation}
\Gamma^{ac}_{bd} (x,y, \; x^\prime , y^\prime ) =
\frac{\delta^4 {\cal S}_I }{\delta q^a(x) \;
\delta q^\dagger_b (x^\prime) \; \delta Q^c(y)
\; \delta Q^\dagger_d  (y^\prime) }
\end{equation}
which is essentially non-local.
In the long-wavelength approximation
the kernel (\ref{22}) factorizes into
two independent kernels as $x \rar x^\prime$ and $y \rar y^\prime$.
The corresponding effective action reads
\beq
{\cal S}_I = - i n N_c \int d^4 z_I \; \int &dU_I&
\bigg [ \; \int d^4x \; {\bf q}^\dagger (x) \; U_I \;
\langle x | \, L_I \, | x \rangle \; U_I^\dagger \; {\bf q}(x) \non \\
&\times&
\int d^4y \; {\bf Q}^\dagger (y) \; U_I \;
\langle y | \, H_I \, | y \rangle \; U_I^\dagger \; {\bf Q}(y) \; \bigg ]
+ I \rar \overline{I}
\enq
and yields the effective interaction in Euclidean space
(leading order in $1/N_c$)

\beq
{\cal L}^E_{qQ} = n \bigg (- \frac {16\pi\rho^3 I_Q}{N_c} \bigg )
\bigg (\frac{4\pi^2\rho^2}{\sqrt{n}\Sigma_0} \bigg ) \left (
i{\bf Q}^\dagger \frac{1+\gamma_4}2 {\bf Q}\,\,i{\bf q}^\dagger {\bf q} +
\frac{1}{4} i{\bf Q}^\dagger \frac{1+\gamma_4}2 \lambda^a {\bf Q}\,\,
i{\bf q}^\dagger  \lambda^a {\bf q} \right )
\label{24}
\enq
For the  detailed construction of the above lagrangian we refer
to the Appendix and our previous paper (\cite{cnz}).
The first bracket in (\ref{24}) arises from the heavy quark part and the second
bracket from the light quark part. Wick-rotating to Minkowski space gives
\beq
{\cal L}_{qQ} = -\bigg (\frac {\Delta M_Q\Delta M_q}{2nN_c} \bigg )\,\,
\left ( \overline{{\bf Q}} \frac{1+\gamma^0}2 {\bf Q}\,\,
\overline{{\bf q}} {\bf q} +
\frac{1}{4} \overline{{\bf Q}} \frac{1+\gamma^0}2 \lambda^a {\bf Q}\,\,
\overline{{\bf q}} \lambda^a {\bf q} \right )
\label{25}
\enq
which is to be compared with the 't Hooft vertex for two light
flavors ${\bf q}=({\bf u}, {\bf d})$
\beq
{\cal L}_{qq} = \bigg ( \frac {\Delta M^2_q}{nN_c} \bigg ) \,\,
\left( {\rm det }\,\overline {\bf q}_R {\bf q}_L \,\,+\,\,{\rm det}\,
\overline {\bf q}_L {\bf q}_R\right)
\label{26}
\enq
Interaction \rf{25} is dominated by the Coulomb-like second term and
has a proper heavy quark spin symmetry. The recoil effect renormalizes the
strength of the interaction through
$\Delta M_Q = \Delta_0 M_Q + \Delta_1 M_Q \sim 86$MeV.
The spin part gives rise to a chromomagnetic interaction

\beq
{\cal L}_{qQ}^{spin} =  \frac{ \Delta M_q \; \Delta M_Q^{spin} }{2nN_c}
\; \frac 14 \;
\overline{\bf Q} \frac{1+\gamma^0}2 \lambda^a \sigma^{\mu\nu} {\bf Q}\,\,
\overline{\bf q} \lambda^a \sigma^{\mu\nu} {\bf q}
\enq
with
\beq
\Delta M_Q^{spin} &=& \frac 1{\rho m_Q}  \; n \; 16 \pi \rho^3
\; \int x^2 \; dx \frac {\sin^2 f_x }{(1+x^2)^2} \simeq 3 \; {\rm MeV}
\enq
for a $c-$quark.
As a Coulomb-like term in \rf{25}, it has a smooth $N_c^0$ limit for
the large $N_c$ and is attractive in the spin zero, color-singlet channel.

Similar
arguments may be applied to the heavy mesons $\overline{Q} Q$  as well. To
order $1/m_Q$ and in the planar approximation, the effective interaction
among the heavy quarks is given by

\beq
{\cal L}_{QQ} = - \bigg (
\frac{ \Delta M_Q \Delta M_Q }{2nN_c} \bigg )
\left ( \overline{{\bf Q}} \frac{1+\gamma^0}2 {\bf Q}\,\,
\overline{{\bf Q}} \frac{1+\gamma^0}2 {\bf Q} +
\frac{1}{4} \overline{{\bf Q}} \frac{1+\gamma^0}2 \lambda^a {\bf Q}\,\,
\overline{{\bf Q}} \frac{1+\gamma^0}2 \lambda^a {\bf Q}
\right ) \label{27}
\enq
The recoil effects appear to first order in $1/m_Q$ and renormalize
$\Delta M_Q$. The spin effects are of second order in $1/m_Q$, and result in

\beq
\Delta {\cal L}_{QQ}^{spin} = \bigg (
\frac{ \Delta M_Q^{spin} \; \Delta M_Q^{spin} }{2nN_c} \bigg ) \;
\frac{1}{4} \overline{{\bf Q}} \frac{1+\gamma^0}2 \lambda^a
\sigma_1^{\mu\nu} {\bf Q}\,\,
\overline{{\bf Q}} \frac{1+\gamma^0}2 \lambda^a \sigma_2^{\mu\nu} {\bf Q}
\label{28}
\enq

For heavy baryons of the type $qqQ$ we have

\beq
{\cal L}_{qqQ} = -\bigg ( \frac{\Delta M_Q \Delta M_q^2}{2n^2 N^2_c} \bigg )
\bigg ( &&\overline{\bf Q} \frac {1+\gamma^0}2 {\bf Q} \,\,
\left({\rm det} \overline {\bf q}_L {\bf q}_R \,\,+{\rm det}\overline
{\bf q}_R {\bf q}_L \,\,\right) +\nonumber\\ &&
       \frac 14 \,\,\overline{\bf Q} \frac {1+\gamma^0}2 \lambda^a {\bf Q} \,\,
\left({\rm det} \overline {\bf q}_L\lambda^a {\bf q}_R \,\,+{\rm det}\overline
{\bf q}_R\lambda^a {\bf q}_L \,\,\right) \bigg )
\label{29}
\enq
and to second order in $1/m_Q$

\beq
{\cal L}_{qqQ}^1 = - \bigg ( \frac{ \Delta M_Q^{spin} \; \Delta
M_q^2}{n^2 N^2_c} \bigg )
\frac 14 \,\,\overline{\bf Q} \frac {1+\gamma^0}2 \lambda^a
\sigma_{\mu\nu} {\bf Q} \;
\bigg ( \; {\rm det} \overline {\bf q}_L\lambda^a
\sigma_{\mu\nu} {\bf q}_R \,\,+{\rm det}\overline
{\bf q}_R\lambda^a \sigma_{\mu\nu} {\bf q}_L \; \bigg ) \label{30}
\enq
The phenomenological implications of these interactions on heavy-light spectra
will be discussed next.

\section{Heavy Hadron Spectra} \label{ch3}

The contributions of the various instanton interactions derived above to the
heavy hadron spectra, will be discussed using a variational approach. For
Mesons, the generic Hamiltonian is

\beq
H = \frac {\vec{p}_q^2}{2m_q}\, +
\frac {\vec{p}_Q^2}{2m_Q}\, +
\frac 12 M \omega^2 |\vec{r}_q - \vec{r}_Q|^2
+ H^{(2)}  \label{harmon}
\enq
where $M$
is the reduced mass of the
heavy-light system, with $m_q = \Delta M_q \sim 420$ MeV and
$m_Q = m_c + \Delta M_Q \sim (1350 + 86)\, $MeV.
The harmonic potential provides for a simple mechanism of confinement.
The instanton-induced interaction $H^{(2)}$, derived from
Eqs. (\ref{25}-\ref{29}) will be treated as a perturbation.
The trial wavefunction is
\beq
\psi (\chi) = \left ( \frac {2\alpha} {\pi} \right )^{3/4}
e^{-\alpha \chi^2 } \label{wf}
\enq
where $\vec\chi = \frac 1{\sqrt 2} (\vec r_q-\vec r_Q)$.
Minimizing the expectation value of \rf{harmon}
in \rf{wf} with respect to $\alpha$ yields $\alpha = \frac 12 M \omega $
with the confining energy ${\cal E}_\alpha = \frac 32 \omega$
as expected. Since the size $r = \sqrt{\frac 1 {2\alpha} } $ of the
ground state is a function of the reduced mass $M$, we fix our parameters
by the size of the heavy-light system $r_{qQ} =0.6\, $fm. Then the size of the
the heavy-heavy system is $r_{QQ} \simeq 0.4\, $fm, and the confining
energy is about ${\cal E}_\alpha \simeq 250$MeV for both of them.

\subsection{Mesons}
For heavy mesons the relevant two-body instanton-induced interactions are
for $Q\overline q$

\beq
H^{(2)}_{qQ} = \bigg (\frac {\Delta M_Q\Delta M_q}{2nN_c} \bigg )\;
\left ( 1 + \frac{1}{4}  \lambda_q^a \lambda_Q^a \right )
\delta ^3 (\vec{r}_q - \vec{r}_Q)  \label{hl2}
\enq
For $Q\overline Q$

\beq
H^{(2)}_{QQ} = \bigg (\frac {\Delta M_Q\Delta M_Q}{2nN_c} \bigg )\;
\left ( 1 + \frac{1}{4}  \lambda_{Q_1}^a \lambda_{Q_2}^a \right )
\delta ^3 (\vec{r}_{Q_1} - \vec{r}_{Q_2})  \label{hh2}
\enq
The induced spin-interaction in the $Q\overline q$ configuration
is given by

\beq
H^{2,s}_{qQ} = - \bigg (\frac {\Delta M_Q^{spin} \Delta M_q}{2nN_c} \bigg )
\, \frac{1}{4} \, {\vec \sigma}_q \cdot {\vec \sigma}_Q \;
\lambda_q^a \cdot \lambda_Q^a \;
\delta ^3 (\vec{r}_q - \vec{r}_Q ) \label{hls}
\enq
We recall that $\Delta M_Q^{spin}$ is down by one power of $1/m_Q$.
Using (\ref{hl2}) and the trial wavefunction (\ref{wf}) we have

\beq
\langle   H^{(2)}_{qQ} \rangle
\sim  - C_F \bigg (\frac {\Delta M_Q\Delta M_q}{2nN_c} \bigg )
\; |\psi(\vec 0 )|^2
= - \frac{N_c}2 \bigg (\frac {\Delta M_Q\Delta M_q}{2nN_c} \bigg )
\; \left ( \frac 1 { \sqrt{\pi} r_{qQ} } \right )^{3}
\enq
and similarly for \rf{hh2}. Thus
$\langle \; H^{(2)}_{qQ} \; \rangle
\sim - 183 \; {\rm MeV}$ and
$\langle \; H^{(2)}_{QQ} \; \rangle
\sim - 103 \; {\rm MeV}$.
These numbers should be compared respectively to $- 140$ MeV and
$- 70$ MeV, as quoted in \cite{cnz} using qualitative arguments.
The spin corrections are of order
$\langle \; H_{qQ}^{2,s} \; \rangle
\simeq - 42 \; {\rm MeV}$, a result that is
consistent with the constituent Quark Model estimate of
$\sim - 27\, $MeV \cite{deruj}. The spin induced interaction in heavy systems
such as charmonium is tiny,
$\langle H_{QQ}^{2,s} \rangle \sim - 0.7\, $MeV.

To evaluate the spectrum in heavy-light and heavy-heavy systems, we use
the mass formulae

\beq
M_{qQ} = \langle  H^{(0)}  + H^{(1)}  + H^{(2)} \rangle
\enq
$H^{(0)}$  is the sum
of the binding energy ${\cal E}_\alpha$ and the current masses
$m_s = 150$MeV,  $m_c = 1350$MeV, $m_b = 4700$MeV, (we take all light
flavors to be massless). $H^{(1)}$ stands for the induced instanton mass
for the light $\Delta M_q \sim 420$MeV and heavy $\Delta M_Q \sim 86$MeV
quarks.   $H^{(2)}$ provides for the extra Coulomb binding energy
discussed above the $1/m_Q$ hyperfine splitting within the multiplets.
Our results are summarized in Table 1. Overall, the results are in reasonable
agreement with experiment and the constituent quark model \cite{deruj}.

\subsection{Baryons} \label{barsp}

Heavy baryons may be analyzed in similar fashion using the induced interactions
discussed at the end of section 3. First, we note that the analog of the
one-gluon exchange in our case, is the induced two-body interaction
(\ref{25}), that is

\beq
H^{(2)}_{qq} = \bigg (\frac {\Delta M_q\Delta M_q}{nN_c} \bigg )\,\,
\left ( 1 + \frac{3}{32}  \lambda_1^a \cdot \lambda_2^a -
\frac{9}{32}  {\bf \sigma}_1 \cdot {\bf \sigma}_2 \;
\lambda_1^a \cdot \lambda_2^a \right )
\delta ^3 (\vec{r}_1 - \vec{r}_2)  \label{ll2}
\enq
and similarly for baryons with two heavy quarks. This interaction is
expected to be overall attractive, thus binding. We note that it scales
like $N_c$ in baryonic configurations. Instantons in heavy baryons, induce
also a three-body interaction (\ref{29}), that is

 \beq
H_{qqQ}^{(3)} &=& -
\bigg ( \frac{\Delta M_Q \Delta M_q^2}{2n^2 N^2_c} \bigg )
\; \sum_{i<j} \delta ^3 (\vec{r}_i - \vec{r}_j ) \;
\delta ^3 (\vec{r}_j - \vec{r}_Q ) \non \\
&\times &
\left\{ {\bf 1}_Q \cdot
\big [ 1 + \frac{3}{32}  \lambda_i^a \cdot \lambda_j^a -
\frac{9}{32}  {\bf \sigma}_i \cdot {\bf \sigma}_j
\lambda_i^a \cdot \lambda_j^a \; \big ] \right. \non \\
&+& \left .  \frac 14 \; \lambda^a_Q \cdot
\big [ \lambda^a_j + \frac{3}{32}  \lambda_i^b \cdot (\lambda^b \lambda^a)_j -
\frac{9}{32}  {\bf \sigma}_i \cdot {\bf \sigma}_j \;
\lambda_i^b \cdot (\lambda^b \lambda^a)_j  + (i\ral j)
\; \big ] \; \right \}
\enq
This interaction scales as $N_c^0$. Although subleading in our previous
book-keeping arguments, we will keep it in our $N_c=3$ arguments.

For baryons, the trial wavefunctions, will be chosen in the form

\beq
\psi (\chi, \eta ) =
\left ( \frac {2 \alpha}{\pi} \right )^{3/2}
e^{-\alpha (\chi^2 +\eta^2)}
\enq
where $\vec\chi = \frac 1{\sqrt 2} (\vec r_1-\vec r_2)$ and
$\vec \eta = \sqrt{\frac 16}  (\vec r_1 + \vec r_2 - 2\vec r_Q )$
are standard Jacobi coordinates.
Here we choose $r_{qqQ}=1\, $fm for the size of the heavy-light baryons.
The size of the heavier configurations will be set to
$r_{qQQ} \simeq 0.86\, $ fm for $Qqq$ and
$r_{QQQ} \simeq 0.7\, $ fm for $QQQ$. The confining energy is
about ${\cal E}_\alpha \sim 500$ MeV for all of them.

The addition of one more quark in the baryonic configurations, brings about in
the three-body contribution to the energy  an additional overall
factor of ${\cal R}_q$ for a light quark, and ${\cal R}_Q$ for a
heavy quark, in comparison to (\ref{ll2}). Specifically,

\beqn
{\cal R}_{q,Q} =  - 2\bigg ( \frac{\Delta M_{q,Q}}{2n N_c} \bigg )
\left ( \frac 1 {\sqrt{\pi} r} \right )^3
\enqn
The three-body contribution is repulsive, whereas  the two-body contribution
 is
attractive. Thus, there is a subtle interplay between two- and
three-body interactions in the determination of the overall energy of the
heavy-light baryonic systems.
For a baryon size $r=1$ fm, ${\cal R}_q\sim -0.75$ and ${\cal R}_Q \sim -0.06$.
We note that the results are very sensitive to the size of the hadron.
Indeed, for $r=0.9$ fm, ${\cal R}_q\sim -1$, and for $r=0.4$ fm,
${\cal R}_Q\sim -1$. In other words, two- and three-body contributions, become
comparable in strength.
The size $r$ is fixed by the choice of the potential (\ref{ll2}) and is
independent from the character of the induced interaction in our discussion.
Here, we chose to work with $r_{qQ}=0.6$ fm
for the heavy-light mesons, and $r_{qqQ}\sim 1$ fm, for the heavy baryons.
A smaller size, say $r_{qqQ}\sim 0.5$ fm would not fit the spectrum.
Of course, other choices may also be possible, with other choices
of the potential in (\ref{harmon}).

We present the spectra for heavy baryons in Table 2.
Our results are in a reasonable agreement with experiment and
other models. Major uncertainty comes from the $1/N_c$ corrections
(non-planar graphs) and the approximation for the ground-state wavefunctions.
As we have seen, the first order corrections were up to 25\% of the
leading order. Since we are making two expansions, $\Lambda_{\rm QCD}/m_Q$
and $1/N_c$, one would also expect substantial contributions
coming from the first order corrections in $1/N_c$ for physical $N_c=3$.

\subsection{Exotics} \label{exosp}

The rationale of constructing two- and three-body interactions using
instanton induced effects, can be extended to multi-quark configurations.
For example, for $QQqq$ configurations, the integration over group
({\em cf.} (\ref{A8})) leads to the  four-body interaction  of the
form

\beq
{\cal L}_{qqQQ} = &&-
n N_c \bigg ( \frac{\Delta M_q}{n N_c} \bigg )^2
\bigg ( \frac{\Delta M_Q} {2 n N_c} \bigg )^2 \non \\
&&\times\bigg (  \overline{\bf Q} \frac {1+\gamma^0}2 {\bf Q} \,\,
\overline{\bf Q} \frac {1+\gamma^0}2 {\bf Q} \,\,
\left({\rm det} \overline {\bf q}_L {\bf q}_R \,\,+{\rm det}\overline
{\bf q}_R {\bf q}_L \,\,\right)  \non \\
&& +\frac 14 \,\,\overline{\bf Q} \frac {1+\gamma^0}2 \lambda^a {\bf Q} \,\,
\overline{\bf Q} \frac {1+\gamma^0}2 {\bf Q} \,\,
\left({\rm det} \overline {\bf q}_L\lambda^a {\bf q}_R \,\,+{\rm det}\overline
{\bf q}_R\lambda^a {\bf q}_L \,\,\right)  \non \\
&& +\frac 14 \,\,\overline{\bf Q} \frac {1+\gamma^0}2 \lambda^a {\bf Q} \,\,
\,\,\overline{\bf Q} \frac {1+\gamma^0}2 \lambda^a {\bf Q} \,\,
\left({\rm det} \overline {\bf q}_L {\bf q}_R \,\,+{\rm det}\overline
{\bf q}_R {\bf q}_L \,\,\right) \bigg )
\enq
The overall sign is consistent with the naive expectation, that the $n$-body
interaction follows from the $(n+1)$-body interaction by contracting a light
quark line, resulting in an overall minus sign (quark condensate).

We recall that for each extra light quark, the penalty factor in the energy
is ${\cal R}_q$. Starting with $r=1$ fm for
three quark states (whether heavy or light), we find that the radius $r$
shrinks to $0.93$ fm for one additional light quark
(four-quark state), to $0.88$ fm for an extra one (five-quark state), and
to $0.84$ fm for still another one (six-quark state). For $r=1$ fm, ${\cal
R}_q\sim -0.72$, while for $r=0.9$ fm, ${\cal R}_q =- 1$. It follows that the
three-body interaction (repulsive) will tend to overcome the binding
energy provided by the two-body interaction (assumed attractive)
in the multiquark configurations of the type
($\bar q \bar q qq$),
($\bar Q \bar q qq$), ($\bar q q \; q qq$), ($\bar Q q \; q qq$) and
($qqq\; qqq$). In this respect, we agree with the conclusions of \cite{oka}
that the H-dibaryon viewed as a six-light-quark state
($qqq\,\, qqq$), will be unbound by the three body-forces induced by instantons
\footnote{In Ref. \cite{oka} the radius $r=0.5$ fm was used in comparison
to $r=0.84$ fm in our case.}.

Adding a heavy quark brings about a penalty factor ${\cal R}_Q$ in the energy.
This factor is 0.06 for $r=1$ fm, and 1 for $r=0.4$ fm. Using the harmonic
potential with two light quarks and four heavy quarks, yield
$r=0.8$ fm and ${\cal R}_Q =0.1$. Thus, for heavy six-quark states
the three-body interaction is 10\% of the two-body interaction, hence small.
In this respect, if we were to think about the H-dibaryon as a
six-heavy-light-quark state ($Qqq\,\,Qqq$) will $not$ be unbound by the
three-body interaction.
Similarly, the four-body-interaction will be expected to be about 1\% of the
two-body, about the same order of magnitude as the hyperfine splitting
discussed above. Therefore we conclude, that the multi-body effects are only
important for multi-quark states near threshold.

We have run specific calculations for exotics containing two and three
heavy quarks of the type ($\bar Q\bar Q \, q q$) and ($\bar Q Q\, Q qq$).
With our choice of parameters, we have found that these configurations
were stable against strong decays through
$\bar Q \bar Q \, q q \rar \bar Q q + \bar Q q $ and
$\bar Q Q\, Q qq \rar \bar Q \bar Q + Q q q$
or $\bar Q Q\, Q qq \rar \bar Q q + Q Q q$.
In both cases the binding energy was found to be of the order of
$10$ MeV, in agreement with other models \cite{sco} and \cite{man}.

\section{Conclusions}

We have presented a general framework for discussing the effects of a dilute
and random instanton gas, on the multiquark configurations involving heavy and
light flavors. Our instanton-induced interactions obey chiral and heavy quark
symmetry to leading order in the bare heavy quark mass. Recoil and
 spin effects
were explicitly worked out and found to be small on single quarks.

We have used the instanton-induced interactions to analyze the spectra of
heavy-light mesons and heavy-light baryons. The results are in overall
agreement with the constituent quark model results as well as soliton
calculations. The role of the three-body force in multiquark states
was also discussed. If the strangeness is viewed as a light degree of freedom,
then the H-dibaryon may be unbound by three-body effects.
 If on the other hand,
strangeness is viewed as a heavy degree of freedom, then the faith of the
H-dibaryon is controlled by the strength of the two-body forces.

We have analyzed the role of multi-body induced instanton interactions on
multi-quark states and found them to be very sensitive to the size of the
states in light systems. The size of the system is fixed by long range
confining forces, and thus outside the scope of the instanton-based models.
In heavy systems, the two-body interaction is
 dominant whatever the size of the
systems considered.

Clearly the present analysis could be extended in several directions. First,
the derivation was based on the planar approximation, and that could be lifted
as subdominant effects may be considered. Second,
 the spectrum calculations may
be refined, by considering more realistic potentials and trial
wavefunctions. A more thorough
analysis of the $1/m_Q$ corrections could be carried along the lines we have
discussed using Bethe-Salpeter construction. In this respect, it would be
interesting to re-investigate directly the present effects on the various
correlation functions. Finally, one could use it to calculate magnetic
moments and other static characteristics of the hadrons.

\acknowledgements
      This work was supported in part by the Department of Energy
under Grant No.\, DE-FG02-88ER40388 and by Grant No.\,2P03B19609
from the Polish
Government Project KBN.

\newpage
\appendix
\renewcommand{\theequation}{A-\arabic{equation}}
\setcounter{equation}{0}
\section{U - Integration}

Here we present a graphical shortcut to the derivation
of the higher order interactions. Our task is to
average over the string of color matrices
\beq
\int dU \, \prod_{k=1}^n \,U_{i_k}^{a_k} U^{\dagger j_k}_{b_k}
\equiv \int dU \; [ \; U_i^a {U^\dagger}_b^j \; ] ^n
\enq
with invariant measure $\int dU = 1$.
Averaging over color is equivalent to finding all projections
onto the singlets of the group, \ie
$P({\bf 3}\otimes \overline{\bf 3})^n \rar {\bf 1}$.
For small $n=1,2,3 $ the answer can be obtained
by a direct calculation using the  method developed by Creutz
\cite{cre}
for averaging over links in lattice gauge theory.
For  higher $n$ the color integration
using this approach becomes very involved and leads to cumbersome
expressions involving sums of $n!^2$ strings of $2n$ Kronecker delta's.
On the other side, since ${\bf 3}\otimes \overline{\bf 3} =
{\bf 1} \oplus {\bf 8}$, the problem
reduces in practice to finding all projections of the product of $n$
octets (adjoint representations) onto the singlet, \ie
$P(\otimes {\bf 8})^n \rar {\bf 1}$ for $SU(N)$ with $N=3$.
The number of distinct projections $A_n$ grows with $n$ like
\beq
 A_n=n!\sum_{k=0}^n (-1)^k \frac{1}{k!}
\label{numerek}
\enq
In order to avoid explicit presentation of the indices, we use the
diagrammatic technique, originally proposed by Cvitanovic \cite{cvi}.

Fundamental graph consists of links and vertices  based on the
following identification (see Fig. 1):
\beq
U_i^a {U^\dagger}_b^j \ral \frac{1}{N} \delta_b^a\; \delta_i^j +
\li_b^a \; \li_i^j
\enq
where
$\lambda^{a}$ $(a=1,\ldots,N^2-1)$ are the color Gell-Mann matrices,
normalized as ${\rm Tr} \lambda^i \lambda^j = 2 \, \delta^{ij}$.

To each projector and symbol we assign a weight as follows
\beq
P_1 = \frac{1}{N} , \qquad P_8 = \frac{1}{4(N^2-1)}
\enq

For $n=1$ one has
\beq
\int dU U_i^a {U^\dagger}_b^j = P({\bf1}\oplus {\bf 8}) = P_1 =
\frac{1}{N} \delta_b^a\; \delta_i^j.
\enq
since averaging over one octet leads to zero.
The indices on the r.h.s. follow from the graphical representation.
$n=2$ makes use of $P_8$, \ie projector of two octets onto singlet  (Fig. 2)
\beq
\int dU [U_i^a {U^\dagger}_b^j]^2 &=
& P({\bf 1}\otimes {\bf 1} + {\bf 8} \otimes {\bf 8})= P_1 \cdot P_1
 + P_8({\bf 8} \otimes {\bf 8}) \non \\
&=&
[ \frac{1}{N} \delta_b^a\; \delta_i^j ]^2 + \frac{1}{4(N^2-1)}
\li_{b_1}^{a_1} \; \li_{b_2}^{a_2} \; \lk_{i_1}^{j_1} \; \lk_{i_2}^{j_2}
\enq
Contributions to the spin $(ij)$ and color $(ab)$ parts are totally
identical, and to simplify the notation, we retain only
the color matrices in our formulae and graphs.

For $n=3$ the only new ingredient is the projection of the product of three
octets onto singlets. Fig. 3 shows how  to project $n=3$.
Explicitly \cite{now},
\beq
P(\otimes {\bf} 8)^3 &=& P[{\bf 8}\otimes({\bf
1}+{\bf 8_a} +{\bf 8_s} +\overline{\bf 10} + {\bf 10} +{\bf 27})] =
P_8[{\bf 8} \otimes {\bf 8_a} + {\bf 8}\otimes {\bf 8_s}] \non \\
&=& \, P_8 P_d \, d({\rm color\cdot spin})
+ P_8 P_f \, f({\rm color\cdot spin}) \non \\
&=& \, \frac{1}{4(N^2-1)} \, \frac{N}{2(N^2-4)} \,
d^{ijk} (\lambda^i_1 \lambda^j_2 \lambda^k_3)_b^a \cdot d({\rm spin})
\non \\
&+& \, \frac{1}{4(N^2-1)} \, \frac{1}{2N} \,
f^{ijk} (\lambda^i_1 \lambda^j_2 \lambda^k_3)_b^a \cdot f({\rm spin})
\enq
where $d({\rm spin}) = d^{abc} (\lambda^a_1 \lambda^b_2 \lambda^c_3)_i^j $
and we have used values of the symmetric and antisymmetric octet projectors
\beq
 P_f = \frac{1}{2N}, \qquad
 P_d = \frac{N}{2(N^2-4)}
\enq
Generalization of the
above procedure for arbitrary $n$ is possible due to the
classification of all invariant tensors for $SU(3)$
by Dittner \cite{dit}. All of them could be constructed from the combinations
of $f^{abd}$ and $d^{abc}$,  standard structure constants defined by
commutator  $[\lambda^a,\lambda^b]_-=2if^{abc} \lambda^c$ and
 anticommutator   $[\lambda^a,\lambda^b]_+= 2 d^{abc} \lambda^c
+ \frac{4}{3}\delta^{ab}$, respectively.
 Assigning a weight for  each symbol: for $f^{abc} \leftrightarrow P_f$,
 and for $d^{abc} \leftrightarrow P_d$, we can derive
the result for arbitrary n
 from the graphical representation.

For $n=4$ one can either have two $P_8$ projections or start with $P_8$
and then contract using four combinations $dd$, $fd$, $df$,
and $ff$
\beq
\int dU [U_i^a {U^\dagger}_b^j]^4 &=&
\frac{1}{N} [\delta_4]_b^a [\delta_4]_i^j \int dU [U_i^a {U^\dagger}_b^j]^3
+ {\rm sym. \; perm's} \non \\
&+& ({\rm spin})_i^j \cdot (\lambda^i_1 \lambda^j_2
\lambda^k_3 \lambda^l_4)_b^a \left \{
\frac{1}{4(N^2-1)} \frac{1}{4(N^2-1)} \left [ \delta^{ij} \delta^{kl} +
\delta^{ik}\delta^{jl} + \delta^{il}\delta^{jk} \right] \right. \non \\
&+& \frac{1}{4(N^2-1)} P_d P_d
\left [ d^{ijm}d^{klm} + d^{ikm}d^{jlm} + d^{ilm}d^{jkm} \right ]
+ d \ral f.
\label{A8}
\enq

Our last example is $n=5$, Fig. 5. Symbolically, it reads
\beq
P(\otimes8)^5 = \bigg [ \, P_8 \cdot P_8 P_d + P_8 \cdot P_d P_d P_d
+ (d \ral f) \, \bigg ] \times ({\rm color \cdot spin})
\enq
Any other $n$ can be considered in a similar
 fashion, and since the explicit
  expression becomes  lengthy, the use of the diagrams
helps greatly in analyzing the properties of the derived formula.

One should keep in mind that not all of possible $(f,d)$
combinations are linearly
independent and using appropriate relations between them \cite{dit}, one
can reduce their number to the number (\ref{numerek}).
Our expression have an advantage of being explicitly symmetric.
  We have tested the procedure in various ways.
  One way is to take  all possible contractions.
For example, as a consistency check,
one may contract a pair of indices, \eg $(b_1a^2)$
$(i_1j^2)$ and see if it is reduced to the $n-1$ case
\beq
\delta^{b_1}_{a_2} \delta^{i_1}_{j_2} \int dU [U_i^a {U^\dagger}_b^j]^n
= N \int dU [U_i^a {U^\dagger}_b^j]^{n-1} \label{contr}
\enq
The same procedure can be used to check formulae for any $n$.
Another useful check is the analysis
of the powers of $N$. The leading behaviour  must scale like  $1/N^n$
for any $n$ as $N \rar \infty$. The leading contribution in this limit
has a form of the determinant build of Kronecker deltas, reproducing, in
the instanton model case, 't Hooft's result for an arbitrary number
of light flavors.

\vfill
\newpage

\centerline{\large \bf Figure Captions}

\vskip 1.1cm

{\bf Figure. 1.} Basic relation.
\bigskip

{\bf Figure. 2.} $n=2$ contraction.
\bigskip

{\bf Figure. 3.} $n=3$ contraction.
\bigskip

{\bf Figure. 4.} $n=4$ contraction.
\bigskip

{\bf Figure. 5.} $n=5$ contraction.

\vfill

\newpage

\centerline{\large \bf Table Captions}

\vskip 1.1cm

{\bf Table I.} Mesonic spectrum.

\bigskip

{\bf Table II.} Baryonic spectrum.

\vfill

\newpage

\input FEYNMAN
\narrowtext
%
%
\begin{figure}
\begin{picture}(20000,18000)
\bigphotons
\THICKLINES
\global\Xone=1000 \global\Yone=14000
\global\Xtwo=\Xone \global\Ytwo=\Yone
\linethickness{0.05cm}
\drawline\fermion[\E\REG](\Xone,\Yone)[7000]
\drawarrow[\E\ATBASE](\pmidx,\pmidy)
\global\advance\pfrontx by -1300
\put(\pfrontx,\pfronty){\Large $a$}
\global\advance\pbackx by 800
\put(\pbackx,\pbacky){\Large $c$}
\global\advance\pbacky by -7000
\global\advance\pbackx by -800
\drawline\fermion[\W\REG](\pbackx,\pbacky)[7000]
\drawarrow[\W\ATBASE](\pmidx,\pmidy)
\global\advance\pfrontx by 800
\put(\pfrontx,\pfronty){\Large $d$}
\global\advance\pbackx by -1300
\put(\pbackx,\pbacky){\Large $b$}
%
%
\global\advance\pbacky by 3300
\global\advance\pbackx by 11800
\put(\pbackx,\pbacky){\huge $\Rightarrow$}
\global\advance\pbackx by 4000
\put(\pbackx,\pbacky){\LARGE $\frac{1}{N}$}
\global\advance\pbackx by 13500
\put(\pbackx,\pbacky){\huge $+$}
%
%
\global\advance\Xone by 19000
\drawline\fermion[\E\REG](\Xone,\Yone)[2000]
\global\advance\pfrontx by -1300
\put(\pfrontx,\pfronty){\Large $a$}
\drawline\fermion[\S\REG](\pbackx,\pbacky)[7000]
\drawarrow[\S\ATBASE](\pmidx,\pmidy)
\drawline\fermion[\W\REG](\pbackx,\pbacky)[2000]
\global\advance\pbackx by -1300
\put(\pbackx,\pbacky){\Large $b$}
\global\advance\pbackx by 7300
\drawline\fermion[\W\REG](\pbackx,\pbacky)[2000]
\global\advance\pfrontx by 800
\put(\pfrontx,\pfronty){\Large $d$}
\drawline\fermion[\N\REG](\pbackx,\pbacky)[7000]
\drawarrow[\N\ATBASE](\pmidx,\pmidy)
\drawline\fermion[\E\REG](\pbackx,\pbacky)[2000]
\global\advance\pbackx by 800
\put(\pbackx,\pbacky){\Large $c$}
%
%
\global\advance\Xone by 14000
\drawline\fermion[\E\REG](\Xone,\Yone)[2000]
\global\advance\pfrontx by -1300
\put(\pfrontx,\pfronty){\Large $a$}
\drawline\fermion[\S\REG](\pbackx,\pbacky)[7000]
\put(\pmidx,\pmidy){\circle*{300}}
\thicklines
\stemmed\drawline\gluon[\E\REG](\pmidx,\pmidy)[4]
\linethickness{0.05cm}
\drawline\fermion[\W\REG](\fermionbackx,\fermionbacky)[2000]
\global\advance\pbackx by -1300
\put(\pbackx,\pbacky){\Large $b$}
\put(\gluonbackx,\gluonbacky){\circle*{300}}
\drawline\fermion[\N\REG](\gluonbackx,\gluonbacky)[3500]
\drawline\fermion[\E\REG](\pbackx,\pbacky)[2000]
\global\advance\pbackx by 800
\put(\pbackx,\pbacky){\Large $c$}
\drawline\fermion[\S\REG](\gluonbackx,\gluonbacky)[3500]
\drawline\fermion[\E\REG](\pbackx,\pbacky)[2000]
\global\advance\pbackx by 800
\put(\pbackx,\pbacky){\Large $d$}
\end{picture}
\vspace*{-1.5cm}
\caption{Basic relation}
\end{figure}

\vskip .3cm
%
%
\begin{figure}
\begin{picture}(20000,18000)
\bigphotons
\global\Xone=15000 \global\Yone=14000
\global\Xtwo=\Xone \global\Ytwo=\Yone
\linethickness{0.07cm}
\drawline\fermion[\S\REG](\Xone,\Yone)[2000]
\thicklines
\stemmed\drawline\gluon[\E\FLAT](\pmidx,\pmidy)[4]
\linethickness{0.07cm}
\drawline\fermion[\N\REG](\pbackx,\pbacky)[1000]
\drawline\fermion[\S\REG](\pfrontx,\pfronty)[1000]
\global\advance\pbacky by -3000
\drawline\fermion[\S\REG](\pbackx,\pbacky)[2000]
\thicklines
\stemmed\drawline\gluon[\W\FLAT](\pmidx,\pmidy)[4]
\linethickness{0.07cm}
\drawline\fermion[\N\REG](\pbackx,\pbacky)[1000]
\drawline\fermion[\S\REG](\pfrontx,\pfronty)[1000]
\global\advance\Ytwo by -4000
\global\advance\Xtwo by 7500
\put(\Xtwo,\Ytwo){\huge  $=$}   
\global\advance\Ytwo by -500
\global\advance\Xtwo by -12000
\put(\Xtwo,\Ytwo){\huge $P_8$}
\global\advance\Xone by 12000
\linethickness{0.07cm}
\drawline\fermion[\S\REG](\Xone,\Yone)[2000]
\thicklines
\drawloop\gluon[\SE 3](\pmidx,\pmidy)
\linethickness{0.07cm}
\drawline\fermion[\N\REG](\pbackx,\pbacky)[1000]
\drawline\fermion[\S\REG](\pfrontx,\pfronty)[1000]
\global\advance\pbackx by 6000
\drawline\fermion[\N\REG](\pbackx,\pbacky)[2000]
\thicklines
\drawloop\gluon[\NW 3](\pmidx,\pmidy)
\linethickness{0.07cm}
\drawline\fermion[\N\REG](\pbackx,\pbacky)[1000]
\drawline\fermion[\S\REG](\pbackx,\pbacky)[2000]
\end{picture}
\vspace*{-1.1cm}
\caption{$n=2$ contraction}
\end{figure}

\vskip .3cm
%
%
\begin{figure}
\begin{picture}(20000,18000)
\bigphotons
\global\Xone=10000 \global\Yone=14000
\global\Xtwo=\Xone \global\Ytwo=\Yone
\linethickness{0.07cm}
\drawline\fermion[\S\REG](\Xone,\Yone)[2000]
\thicklines
\stemmed\drawline\gluon[\E\FLAT](\pmidx,\pmidy)[4]
\linethickness{0.07cm}
\drawline\fermion[\N\REG](\pbackx,\pbacky)[1000]
\drawline\fermion[\S\REG](\pfrontx,\pfronty)[1000]
\global\advance\pbacky by -3700
\drawline\fermion[\S\REG](\pbackx,\pbacky)[2000]
\thicklines
\stemmed\drawline\gluon[\W\FLIPPEDFLAT](\pmidx,\pmidy)[4]
\linethickness{0.07cm}
\drawline\fermion[\N\REG](\pbackx,\pbacky)[1000]
\drawline\fermion[\S\REG](\pfrontx,\pfronty)[1000]
\global\advance\pbacky by -3700
\drawline\fermion[\S\REG](\pbackx,\pbacky)[2000]
\thicklines
\stemmed\drawline\gluon[\E\FLIPPEDFLAT](\pmidx,\pmidy)[4]
\linethickness{0.07cm}
\drawline\fermion[\N\REG](\pbackx,\pbacky)[1000]
\drawline\fermion[\S\REG](\pfrontx,\pfronty)[1000]
\global\advance\Ytwo by -7500
\global\advance\Xtwo by -4500
\put(\Xtwo,\Ytwo){\huge $P_8$}
\global\advance\Ytwo by 200
\global\advance\Xtwo by 13000
\put(\Xtwo,\Ytwo){\huge  $=$}   
%
%
\global\advance\Xone by 13000
\linethickness{0.07cm}
\drawline\fermion[\S\REG](\Xone,\Yone)[2000]
\thicklines
\drawline\gluon[\SE\REG](\pmidx,\pmidy)[5]
\put(\pbackx,\pbacky){\circle*{300}}
\put(\pbackx,\pbacky){\circle{800}}
\global\Xthree=\pbackx \global\Ythree=\pbacky
\drawline\fermion[\W\REG](\pbackx,\pbacky)[1000]
\stemmed\drawline\gluon[\W\REG](\pbackx,\pbacky)[4]
\linethickness{0.07cm}
\drawline\fermion[\N\REG](\pbackx,\pbacky)[1000]
\drawline\fermion[\S\REG](\pfrontx,\pfronty)[1000]
\thicklines
\drawline\gluon[\SW\REG](\Xthree,\Ythree)[5]
\linethickness{0.07cm}
\drawline\fermion[\N\REG](\pbackx,\pbacky)[1000]
\drawline\fermion[\S\REG](\pfrontx,\pfronty)[1000]
%
%
\global\advance\gluonfrontx by 3000
\thicklines
\drawline\gluon[\NE\REG](\gluonfrontx,\gluonfronty)[5]
\put(\pfrontx,\pfronty){\circle*{300}}
\put(\pfrontx,\pfronty){\circle{800}}
\global\Xfour=\pfrontx \global\Yfour=\pfronty
\linethickness{0.07cm}
\drawline\fermion[\N\REG](\pbackx,\pbacky)[1000]
\drawline\fermion[\S\REG](\pfrontx,\pfronty)[1000]
\thicklines
\drawline\fermion[\E\REG](\gluonfrontx,\gluonfronty)[1000]
\stemmed\drawline\gluon[\E\REG](\pbackx,\pbacky)[4]
\linethickness{0.07cm}
\drawline\fermion[\N\REG](\pbackx,\pbacky)[1000]
\drawline\fermion[\S\REG](\pfrontx,\pfronty)[1000]
\thicklines
\drawline\gluon[\SE\FLIPPED](\Xfour,\Yfour)[5]
\linethickness{0.07cm}
\drawline\fermion[\N\REG](\pbackx,\pbacky)[1000]
\drawline\fermion[\S\REG](\pfrontx,\pfronty)[1000]
\global\advance\Ytwo by 3800
\global\advance\Xtwo by 8400
\put(\Xtwo,\Ytwo){\huge $(\, d , f\, )$}
\end{picture}
\vspace*{.4cm}
\caption{$n=3$ contraction}
\end{figure}

\vskip 1.2cm
%
%
\begin{figure}
\begin{picture}(20000,18000)
\bigphotons
\global\Xone=8000 \global\Yone=14000
\linethickness{0.07cm}
\drawline\fermion[\S\REG](\Xone,\Yone)[2000]
\thicklines
\drawloop\gluon[\SE 3](\pmidx,\pmidy)
\linethickness{0.07cm}
\drawline\fermion[\N\REG](\pbackx,\pbacky)[1000]
\drawline\fermion[\S\REG](\pfrontx,\pfronty)[1000]
%
%
\global\advance\pbacky by -3000
\drawline\fermion[\S\REG](\pbackx,\pbacky)[2000]
\thicklines
\drawloop\gluon[\SE 3](\pmidx,\pmidy)
\linethickness{0.07cm}
\drawline\fermion[\N\REG](\pbackx,\pbacky)[1000]
\drawline\fermion[\S\REG](\pfrontx,\pfronty)[1000]
\global\advance\pbackx by 6000
\global\advance\pbacky by 7500
\put(\pbackx,\pbacky){\huge $+$}
\global\advance\pbackx by 17000
\put(\pbackx,\pbacky){\huge $+\; $ perm's}
%
%
\global\advance\Xone by 11000
\drawline\fermion[\S\REG](\Xone,\Yone)[2000]
\thicklines
\drawline\gluon[\SE\REG](\pmidx,\pmidy)[4]
\put(\pbackx,\pbacky){\circle*{300}}
\put(\pbackx,\pbacky){\circle{800}}
\drawline\fermion[\W\REG](\pbackx,\pbacky)[1000]
\global\Xtwo=\pfrontx \global\Ytwo=\pfronty
\stemmed\drawline\gluon[\W\FLIPPED](\pbackx,\pbacky)[3]
\linethickness{0.07cm}
\drawline\fermion[\N\REG](\pbackx,\pbacky)[1000]
\drawline\fermion[\S\REG](\pfrontx,\pfronty)[1000]
%
%
\global\advance \gluonfrontx by 2000
\global\advance \gluonfronty by 1000
\put(\gluonfrontx,\gluonfronty){\huge $(\, d , f\, )$}
\thicklines
\drawline\fermion[\S\REG](\Xtwo,\Ytwo)[1000]
\drawline\gluon[\S\REG](\pbackx,\pbacky)[3]
\drawline\fermion[\S\REG](\pbackx,\pbacky)[1000]
\put(\pbackx,\pbacky){\circle*{300}}
\put(\pbackx,\pbacky){\circle{800}}
\global\Xsix=\pbackx \global\Ysix=\pbacky
\drawline\fermion[\W\REG](\pbackx,\pbacky)[1000]
\stemmed\drawline\gluon[\W\REG](\pbackx,\pbacky)[3]
\linethickness{0.07cm}
\drawline\fermion[\N\REG](\pbackx,\pbacky)[1000]
\drawline\fermion[\S\REG](\pfrontx,\pfronty)[1000]
\thicklines
\drawline\gluon[\SW\REG](\Xsix,\Ysix)[4]
\linethickness{0.07cm}
\drawline\fermion[\N\REG](\pbackx,\pbacky)[1000]
\drawline\fermion[\S\REG](\pfrontx,\pfronty)[1000]
%
%
\global\advance \gluonfrontx by 500
\global\advance \gluonfronty by - 2000
\put(\gluonfrontx,\gluonfronty){\huge $(\, d , f\, )$}
%
\end{picture}
\vspace*{1.5cm}
\caption{$n=4$ contraction}
\end{figure}

\vskip 1.5cm
%
%
\begin{figure}
\begin{picture}(20000,18000)
\bigphotons
\global\Xone=2000 \global\Yone=14000
\global\Xtwo=\Xone \global\Ytwo=\Yone
\linethickness{0.07cm}
\drawline\fermion[\S\REG](\Xone,\Yone)[2000]
\thicklines
\drawloop\gluon[\SE 3](\pmidx,\pmidy)
\linethickness{0.07cm}
\drawline\fermion[\N\REG](\pbackx,\pbacky)[1000]
\drawline\fermion[\S\REG](\pfrontx,\pfronty)[1000]
\global\advance\pbacky by -3000
\drawline\fermion[\S\REG](\pbackx,\pbacky)[2000]
\thicklines
\drawline\gluon[\SE\REG](\pmidx,\pmidy)[5]
\put(\pbackx,\pbacky){\circle*{300}}
\put(\pbackx,\pbacky){\circle{800}}
\global\Xfour=\pbackx \global\Yfour=\pbacky
\drawline\fermion[\W\REG](\pbackx,\pbacky)[1000]
\stemmed\drawline\gluon[\W\REG](\pbackx,\pbacky)[4]
\linethickness{0.07cm}
\drawline\fermion[\N\REG](\pbackx,\pbacky)[1000]
\drawline\fermion[\S\REG](\pfrontx,\pfronty)[1000]
\thicklines
\drawline\gluon[\SW\REG](\Xfour,\Yfour)[5]
\linethickness{0.07cm}
\drawline\fermion[\N\REG](\pbackx,\pbacky)[1000]
\drawline\fermion[\S\REG](\pfrontx,\pfronty)[1000]
\global\advance\pbackx by 13000
\global\advance\pbacky by 10500
\put(\pbackx,\pbacky){\huge $+$}
\global\advance\pbackx by 24000
\put(\pbackx,\pbacky){\huge $+\; $ perm's}
%
%
\global\advance \gluonfronty by 1000
\put(\gluonfrontx,\gluonfronty){\huge $(\, d , f\, )$}
%
%
\global\advance\Xone by 17000
\drawline\fermion[\S\REG](\Xone,\Yone)[2000]
\thicklines
\drawline\gluon[\SE\REG](\pmidx,\pmidy)[5]
\put(\pbackx,\pbacky){\circle*{300}}
\put(\pbackx,\pbacky){\circle{800}}
\backstemmed\drawline\gluon[\W\FLIPPEDFLAT](\pbackx,\pbacky)[4]
\linethickness{0.07cm}
\drawline\fermion[\N\REG](\pbackx,\pbacky)[1000]
\drawline\fermion[\S\REG](\pfrontx,\pfronty)[1000]
%
%
\global\advance \gluonfrontx by 500
\global\advance \gluonfronty by 1000
\put(\gluonfrontx,\gluonfronty){\huge $(\, d , f\, )$}
\global\advance \gluonfronty by -1000
\global\advance \gluonfrontx by -500
\thicklines
\drawline\gluon[\SE\REG](\gluonfrontx,\gluonfronty)[5]
\put(\pbackx,\pbacky){\circle*{300}}
\put(\pbackx,\pbacky){\circle{800}}
\global\Xsix=\pbackx \global\Ysix=\pbacky
\drawline\fermion[\W\REG](\pbackx,\pbacky)[1000]
\stemmed\drawline\gluon[\W\REG](\pbackx,\pbacky)[9]
\linethickness{0.07cm}
\drawline\fermion[\N\REG](\pbackx,\pbacky)[1000]
\drawline\fermion[\S\REG](\pfrontx,\pfronty)[1000]
%
%
\global\advance \gluonfrontx by 2000
\put(\gluonfrontx,\gluonfronty){\huge $(\, d , f\, )$}
\global\advance \gluonfrontx by -2000
\thicklines
\drawline\gluon[\SW\REG](\Xsix,\Ysix)[5]
\put(\pbackx,\pbacky){\circle*{300}}
\put(\pbackx,\pbacky){\circle{800}}
\backstemmed\drawline\gluon[\W\REG](\pbackx,\pbacky)[5]
\linethickness{0.07cm}
\drawline\fermion[\N\REG](\pbackx,\pbacky)[1000]
\drawline\fermion[\S\REG](\pfrontx,\pfronty)[1000]
\thicklines
\drawline\gluon[\SW\REG](\gluonfrontx,\gluonfronty)[5]
\linethickness{0.07cm}
\drawline\fermion[\N\REG](\pbackx,\pbacky)[1000]
\drawline\fermion[\S\REG](\pfrontx,\pfronty)[1000]
%
%
\global\advance \gluonfrontx by 500
\global\advance \gluonfronty by - 2000
\put(\gluonfrontx,\gluonfronty){\huge $(\, d , f\, )$}
%
\end{picture}
\vspace*{4.cm}
\caption{$n=5$ contraction}
\end{figure}

\newpage
\narrowtext

\begin{table}
\caption{Mesonic spectrum}
\label{tab:one}
\begin{tabular}{ccccc}
          & $I(J)^P$  & Exp. value & Prediction &
QM \cite{deruj} \\
\hline
$D^0      $ & $\frac 12 (0^-)$ & 1869 & 1881 & 1800 $\div$ 1860 \\
$D^*      $ & $\frac 12 (1^-)$ & 2010 & 1965 & 1930 $\div$ 1990 \\
$D^{\pm}_s$ & $0 (0^-)$        & 1969 & 2031 & 1975 \\
$D^{\pm}_s$ & $0 (1^-)$        & 2110 & 2115 & 2061 \\
\hline
$B^0  $     & $\frac 12 (0^-)$ & 5278 & 5231 &      \\
\end{tabular}
\end{table}

\begin{table}
\caption{Baryonic spectrum}
\label{tab:two}
\begin{tabular}{ccccccc}
  & I& J & Exp. value & Prediction& QM \cite{deruj} & Ref. \cite{sco} \\
\hline
$\Lambda_c$ & 0 & $\frac12$ & (2285) & 2376 & 2200 & 2170 \\
$\Sigma_c$  & 1 & $\frac12$ & (2453) & 2502 & 2360 & 2421 \\
$\Xi_c$ & $\frac 12$ &$ \frac12 $ & 2468  & 2652 & 2420 & 2421 \\
$\Omega_c  $   & 0 & $\frac 12$ & (2704) & 2802 & 2680 & 2645 \\
\hline
$\Xi_{cc}  $&$\frac 12$& $\frac 12 $ & ? & 3558 & 3550 & 3510 \\
$\Omega_{cc} $ & 0 & $\frac 12 $   & ? & 3708 & 3730 & 3698 \\
$\Omega_{ccc}$ & 0 &$ \frac 32  $  & ? & 4808 & 4810 & 4784 \\
\end{tabular}
\end{table}

\end{document}